\documentclass[prl,twocolumn,aps,showpacs,amssymb]{revtex4}
\usepackage{graphicx}
\usepackage{epsfig}
\usepackage{color}
\newcommand{\nn}[1]{{\langle #1 \rangle}}

\newcommand{\hc}{\mathrm{h.c.}}

\newcommand{\Tbkt}{{T_\mathrm{BKT}}}

\begin{document}

\title{Pair Superfluidity of Three-Body Constrained Bosons in Two Dimensions}
\author{Lars Bonnes and Stefan Wessel}
\affiliation{Institut f\"ur Theoretische Physik III,
Universit\"at Stuttgart, Pfaffenwaldring 57, 70550 Stuttgart, Germany}
\begin{abstract}
We examine the equilibrium properties of lattice bosons with attractive on-site interactions in the presence of a three-body hard-core constraint that stabilizes the system against collapse and gives rise to a dimer superfluid phase formed by virtual hopping processes of boson pairs. 
Employing quantum Monte Carlo simulations, the ground state phase 
diagram of this system on the square lattice is analyzed. 
In particular, we study the quantum phase transition between the atomic and dimer superfluid regime and analyze the nature of the superfluid-insulator transitions. 
Evidence is provided for the existence of a tricritical point along the saturation transition line, where the transition changes from being first-order to a continuous transition of the dilute bose gas of holes. 
The Berzinskii-Kosterlitz-Thouless transition from the dimer superfluid to the normal fluid is found to be consistent with an anomalous stiffness jump, as expected from the unbinding of half-vortices. 
\end{abstract}
\maketitle

Due to their remarkable versatility, cold-atom systems are ideally suited to realize
quantum simulators for the physics of strong correlations~\cite{lewenstein07}.
A fascinating approach towards enhancing correlations in atomic systems steams from
the fact that they can emerge via dissipative processes. In fact,  dissipation-induced correlation effects were observed in an experiment with Feshbach molecules subject to strong inelastic collisions~\cite{syassen08}. 
More recently, it was found that three-body loss processes for bosons in an optical lattice give rise to an effective Bose-Hubbard model description with a local three-body hardcore constraint~\cite{daley09}. Such a constraint stabilizes the system in the presence of strong attractive interactions, where dimer bound states proliferate.  
The effective lattice model describing this 
situation is the Bose-Hubbard Hamiltonian
\begin{equation}
H=-t \sum_\nn{ij} \left( b^\dagger_i b_j + \hc \right)
  +\frac{U}{2} \sum_i n_i(n_i-1)
  -\mu \sum_i n_i
\end{equation}
where $t$ denotes the tunneling matrix element for nearest neighbor sites $\nn{ij}$ and $U<0$ an attractive  on-site interaction~\cite{daley09}. The filling $n$ is controlled by varying the chemical potential $\mu$,
$b_i$ ($b_i^\dagger$) denote bosonic annihilation (creation) operators and $n_i$ the number operator for bosons on lattice site $i$. In contrast to the usual Bose-Hubbard model, 
the Hilbert space is now restricted by the contraint $(b_i^\dagger)^3=0$ to a maximum of two bosons on each lattice site.
In another recent proposal~\cite{mazza10}, a similar effective lattice model of bosons with a three-body constraint was derived for spin-1 atoms, which
in addition includes an explicit correlated hopping term $H'=-t' \sum_\nn{ij} \left( b^{\dagger 2}_i b^2_j + \hc \right)$ of  the dimer bound states. 

%
\begin{figure}
\includegraphics[width=\columnwidth]{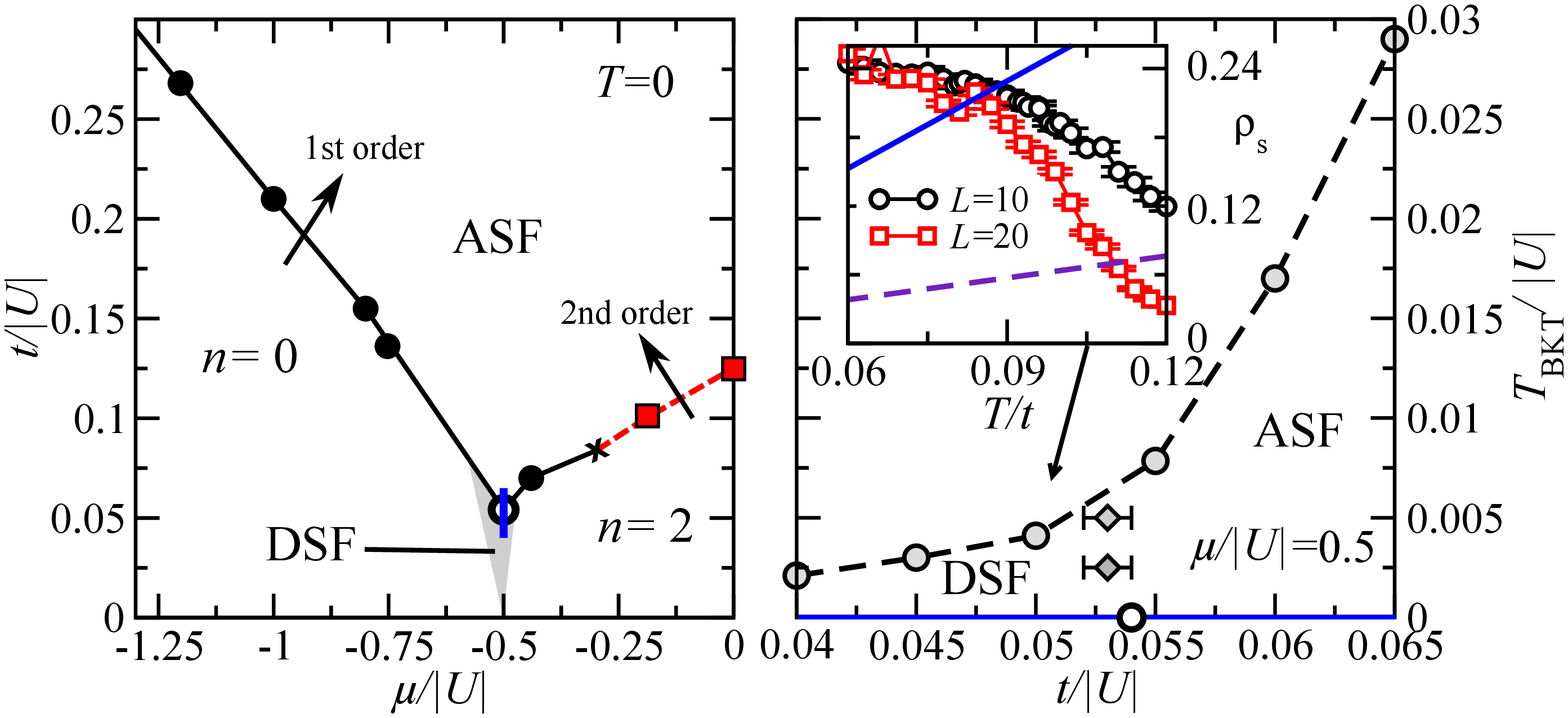}
\label{fig:figure1}
\caption{(Color online) {Left panel:} Ground state phase diagram of the three-body constrained Bose-Hubbard model with attractive on-site interactions on the square lattice. Continuous (first-order) quantum phase transitions are shown by dashed (solid) lines. The cross indicates a tricritical point. The gray area contains the estimated DSF region. {Right panel:} Finite temperate phase diagram along $\mu/|U|=-0.5$. The inset shows the temperature dependence of the superfluid density $\rho_s$ at $t/|U|=0.05$ (DSF) for $L=10$ and $20$ along with lines for  normal $2T/\pi$ (dashed line) and  anomalous $8T/\pi$ (solid line) universal stiffness jumps. 
}
\end{figure}
 
The model in Eq.~(1) exhibits an intriguing phase diagram, shown in Fig. 1 for the case of a two-dimensional square lattice. Recent analytical calculations~\cite{daley09,diehl09a,diehl10,lee10} and numerical works 
for the one-dimension case~\cite{daley09,diehl09a,diehl10} exhibited that besides the trivially insulating phases at $n=0$ and $n=2$, the system stabilizes two kinds of superfluid phases. 
The atomic superfluid (ASF) is characterized by a finite atomic condensate  with $\nn{b_i}\neq 0$ and a finite superfluid response $\rho_s$. For strong interactions $|U|\gg t$ however, a dimer (pair) superfluid phase (DSF)~\cite{valatin58} is stabilized, which is characterized by a vanishing atomic condensate and  $\nn{b_i} = 0$, but a finite dimer condensate density  with an order parameter $\nn{(b_i)^2}\neq 0$. Such
single-component DSF phases have  been observed before in models with explicit correlated or  pair hopping processes~\cite{schmidt06,zhou09,mazza10}. 
In the DSF phase,
the $U(1)$ symmetry of the Hamiltonian is partially broken down to $\mathbb{Z}_2$, and at the DSF to ASF transition, this remaining $\mathbb{Z}_2$ symmetry gets broken. With respect to this partial internal symmetry breaking, DSF are thus related to spin nematic states~\cite{andreev84}. Within Ginzburg-Landau theory, 
a Feshbach resonance term couples the ASF and DSF order parameter fields, which implies an effective $U(1)\times \mathbb{Z}_2$ symmetry similar as in the boson Feshbach resonance problem~\cite{radzihovsky04,radzihovsky08,diehl09a,lee10}.
From such analysis,
the ASF-DSF transition was found to be Ising-like at unit filling, $n=1$, and driven first-order by fluctuations via the  Coleman-Weinberg
mechanism ~\cite{coleman73} for $n\ne1$~\cite{diehl09a,diehl10}. 

Here, we study the ground state and thermal phase diagram of the three-body constrained Bose-Hubbard model in Eq.~(1) using quantum Monte Carlo (QMC) simulations. We focus on a two-dimensional square lattice geometry, on which true off-diagonal long-range order (ODLRO) within the ASF and DSF regimes emerges in the ground state. Besides establishing the presence of ODLRO, we assess the above mentioned theory for the ASF-DSF transition~\cite{diehl09a,diehl10} as well as a recent effective potential 
approach to the insulator-ASF quantum phase transitions~\cite{lee10}, which finds that they are driven first-order close to $n=1$. We indeed obtain evidence for a tricritical point along one of these transition lines.
The thermal Berzinskii-Kosterlitz-Thouless (BKT) transition out of the DSF phase is found to be consistent with an anomalous jump in the superfluid density, driven by the unbinding of half-vortices~\cite{mukerjee06}.

{\it Method}.--
We employ a generalized directed loop algorithm in the stochastic series expansion  (SSE)  representation~\cite{sandvik99b,alet05} at finite temperatures $T$. The simulations are performed on finite square lattices of linear extend $L$ (and $N=L^2$ sites), with periodic boundary conditions, such that
the superfluid density is obtained from the winding number $W$ fluctuations, $\rho_s=T\nn{W^2}/(2t)$~\cite{pollock87,bernardet02}. Since ODLRO in our system is forbidden at finite $T$~\cite{mermin66}, we need to
perform the simulations at sufficiently low $T$ to probe ground state 
properties of the finite system, as detailed below. 
Using two kinds of directed loops, in which the worm heads carry
either a single creation (annihilation) operator $b^\dagger$ ($b$) or a pair operator $(b^\dagger)^2$ ($(b)^2$), provides direct access to the equal-time Green's function of the atoms $G_1(i,j)=\nn{b_i^\dagger b_j}$ as well as the dimers $G_2(i,j)=\nn{ (b_i^\dagger)^2 (b_j)^2}$ \cite{dorneich01,alet05}. The atomic and dimer condensate densities are obtained as $C_1=1/N^2 \sum_{ij} G_1(i,j)$ and $C_2=1/N^2 \sum_{ij}G_2(i,j)$ respectively after extrapolations to the thermodynamic limit. 
A similar scheme with pair-worms was shown recently to be efficient for simulating  two-component boson systems~\cite{ohgoe10}. Here,  we find that
accessing the dimer condensate density based on $G_2$ still  becomes problematic
at the relevant low temperatures: 
Histograms of  
individual measurements  for $C_2$ (related to the lengths of  the pair 
operator loops) exhibit fat-tailed distributions, i.e. the estimator of this quantity is dominated by rare events that make its sampling inefficient. A typical histogram within the DSF region is shown in Fig.~2. 
This behavior results form the fact that pairs of bosons proliferate 
at large $|U|$ near $n=1$. A worm head carrying a bosonic pair operator performs an off-diagonal move (corresponding to the hopping of a dimer) only if it encounters a bond that shares a dimer (or an empty site, if the worm either carries annihilation or creation operators) and a single atom. Such processes are thus strongly suppressed in the relevant parameter regime. 
\begin{figure}
\includegraphics[width=\columnwidth]{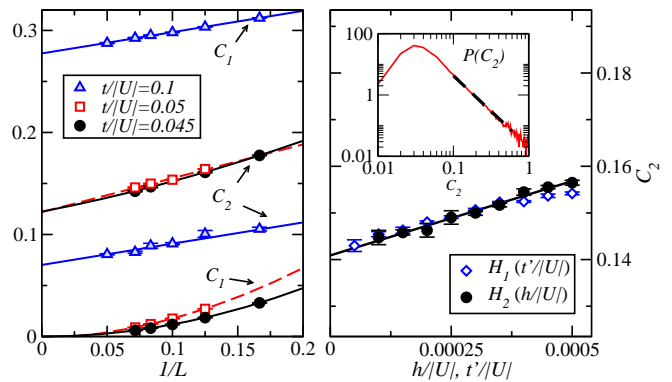}
\label{fig:figure2}
\caption{(Color online)
{Left panel:} Finite size scaling of $C_1$ and $C_2$ for different values of $t/|U|$ from simulations at $T/|U|=0.01/L$.
{Right panel:} Extracting $C_2$ from simulations of $H_1$ and $H_2$. 
Inset: Histogram $P(C_2)$ from simulations of $H$, with a power-law fit $P(C_2)\propto (C_2)^\alpha$, $\alpha=-2.2$ (dashed line) to the fat tail 
at $t/|U|=0.045$, $\mu/|U|=-0.5$, $L=14$ and $T/|U|=0.002$.
}
\end{figure}
Moreover, 
the fat tail of the histogram fits well to a Fr\'{e}chet distribution.
The exponent of the power-law decay in the tail is $\alpha>-3$ ($\alpha\approx -2.2$ for the histogram in Fig. 2), so that the variance does not exist, and the central limit theorem
for the mean value does not hold.

Hence, we resign  to  alternative estimates of the dimer condensate density in the DSF region. 
We employ
two different means of adding correlated hopping terms to the original Hamiltonian that allow for efficient
updates of the boson pairs. The results from both approaches agree within error bars, and with results based on $G_2$ in cases, where the distribution 
of the loop lengths leads to a finite variance. 
In the first approach, we add to $H$ the correlated hopping term $H'$, such that $H_1=H+H'$.  The other approach 
couples $C_2$ directly to the Hamiltonian, such that
$
H_2=H-h/N\sum_{ij} \left( (b_i^\dagger)^2 (b_j)^2 + \hc + \mathbf{1}\right)
$
features correlated hopping terms between all sites of the system (the diagonal term allows the insertion of long-range vertices into the SSE operator string). The coupling $h/N$ ensures an extensive  energy.
Besides the diagonal update of the short-range terms in $H_2$, we 
insert/remove long-range vertices using heat-bath 
probabilities~\cite{sandvik03}. We can now measure 
$C_2$ based on the estimator $C_2= \nn{\sum_b (H_b)^2}/(N t)^2 
-\nn{H_h}/(N h) $, where $H_h$ denotes all  correlated hopping terms in 
$H_2$ and $H_b$ the atomic kinetic energy term on bond $b$. This method 
remains robust  down to very low values of $h/|U|\sim 10^{-4}$, so that  
we  extract the condensate density of the  model $H$ by fitting to a low-degree polynomial (the data is found to be 
essentially linear up to $t'/|U|,~h/|U| \sim 10^{-2}$), 
performing a bootstrap analysis for the error estimation. 
Such a scaling  is shown in the right panel of Fig.~2, and extrapolations 
to the thermodynamic limit in the left panel.

\begin{figure}
\includegraphics[width=\columnwidth]{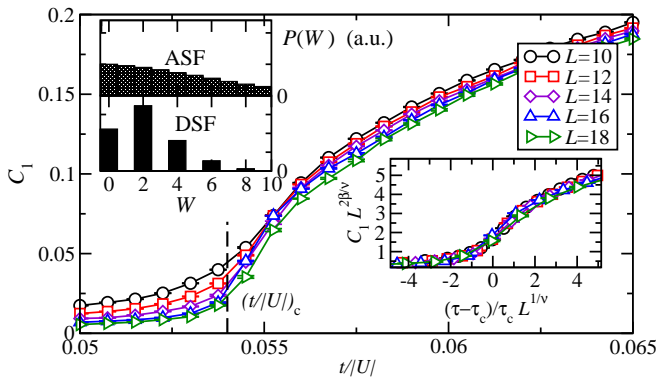}
\label{fig:figure3}
\caption{(Color online)
QMC data for $C_1$ obtained for different system sizes along $\mu/|U|=-0.5$ for  $T/|U|=0.01/L$. 
The critical coupling $(t/|U|)_c \approx 0.054$ is denoted by a vertical bold line.
Right inset: data collapse based on three-dimensional Ising critical exponents. 
Left inset: Histogram $P(W)$ of the winding number $W$ taken in the  ASF ($t/|U|=0.06$) and DSF ($t/|U|=0.045$) region at $T/|U|=0.001$.
}
\end{figure}

{\it ASF and DSF phases}.--
We concentrate on the fixed line $\mu/|U|=-0.5$ (cf. Fig.~1) and vary $t/|U|$ 
in order to reach both the ASF and the DSF phases.  
The finite-temperature phase diagram  along this line (cf. Fig. 1) exhibits the temperature scale at which the system undergoes a BKT transition,
below which superfluidity emerges. The transition temperature $\Tbkt$ was determined from the finite size scaling of $\rho_S$ with system sizes up to $L=30$, following Ref.~\cite{weber87}. As seen from Fig.~1, the slope of the BKT-line changes near $t/|U| \approx0.054$. For  smaller $t/|U|$, the system enters the DSF regime and the superflow is driven by virtual pair hopping processes. Correspondingly, we observe  in the DSF region a strong even-odd effect in the winding number~\cite{schmidt06}  (i.e. only even values are measured), which is absent in the ASF regime, cf. the left inset in Fig.~3.
This allows to locate the $T>0$ phase boundary separating the ASF and DSF phases (diamonds in Fig.~1). Moreover, 
due to the 
nematic nature of the DSF, the BKT transition to this paired phase is driven by  unbinding of half-vortices instead of the usual integer vortices as for e.g. the BKT transition to the ASF phase~\cite{mukerjee06}. This leads to an anomalous universal $8\Tbkt/\pi$ stiffness jump at $\Tbkt$ instead of $2\Tbkt/\pi$ as for the ASF phase~\cite{mukerjee06}. This anomaly is consistent with the finite size behavior of $\rho_S$ shown in the inset of Fig.~1, and was accounted for in the determination of $\Tbkt$ based on the scheme in Ref.~\cite{weber87}.
To assess the extend of the ASF phase within the ground state, we measured  
$C_1$ for different system sizes at $T/|U|=0.01/L$,
which we found necessary in order to access the finite system's  ground state properties. The finite size data is shown in the main panel of Fig. 3.
From  extrapolations of $C_1$ to the thermodynamic limit such as shown in Fig.~2 (left panel), 
we find that below $t/|U|\lesssim0.054$ the atomic condensate density vanishes in the thermodynamic limit. Since 
$\rho_S$  remains finite below this value, we still expect the emergence of ODLRO from a dimer condensate $C_2$ beyond the ASF regime.
In fact, the data of $C_2$ shown in Fig.~2 extrapolates to finite values both within the ASF and the DSF region. 

{\it ASF - DSF quantum phase transition}.--
Upon fixing $\mu/|U|=-0.5$, we cross the ASF-DSF quantum phase transition line slightly away from unit filling $n=1$ (the density for $t/|U|=0.054$ is about $1.04$), so it is expected that the transition at fixed $\mu/|U|=-0.5$ is  
first-order~\cite{diehl09a,diehl10}. However, as pointed out in Refs.~\cite{diehl09a}, the correlation length in the vicinity of the  Ising critical point at $n=1$ is expected to be large, diverging as $(1-n)^{-6}$.
Close to $n=1$, this severely exceeds the finite sizes accessible to our simulations. Based on this scenario, 
the accessible finite systems could be still controlled by the adjacent Ising critical point at $n=1$. 
We indeed observe an approximate scaling in the data of $C_1$ close to the transition point, described by the standard finite-size scaling ansatz
$
C_1(L,\tau)=L^{-2\beta/\nu} g(L^{1/\nu}(\tau-\tau_c)/\tau_c),
$
where $\tau=t/|U|$, and $\tau_c$ denotes the transition point.  
In the inset of Fig. 3, we show a corresponding data collapse of the available finite-system data for $\mu/|U|=-0.5$ from the main panel, using three-dimensional Ising critical exponents $\beta=0.3264(4)$ and $\nu=0.6298(5)$~\cite{hasenbusch99a} and $T/|U|=0.01/L$
(for a dynamical critical exponent $z=1$), giving $(t/|U|)_c=0.054(1)$.
Unfortunately, we are not able to discern whether the quantum phase transition  indeed is first-order or not (histograms  of various observables did not exhibit any two-peak structures on the accessible system sizes $L<20$ at these low temperatures). Controlling $\mu$ as to fix $n=1$ for all finite systems in order to directly address the $n=1$ case also turned out unfeasible.
Given the above estimate of $(t/|U|)_c$,
the mean-field value $(t/|U|)_c\approx 0.044$~\cite{daley09} is found to underestimate the stability of the DSF phase by about $20\%$. The calculations of Ref.~\cite{diehl09a}, while accounting for  quantum fluctuations effects, still show a similar deviation. 

\begin{figure}
\includegraphics[width=8cm]{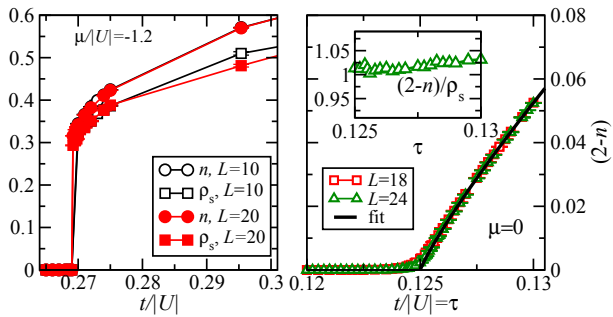}
\label{fig:figure4}
\caption{(Color online)
Left Panel: Density $n$ and superfluid density $\rho_s$ near the ASF-insulator 
transitions at $\mu/|U|=-1.2$ ($T/|U|=0.008$). 
Right Panel: Hole density $(2-n)$ at  $\mu=0$ and $T/|U|=0.08/L$ for various system sizes fitted to $(2-n)=a(\tau-\tau_c) \ln(b/(\tau-\tau_c))$ for the dilute bose gas transition at $\tau_c=0.125$. 
The inset shows the ratio $(2-n)/\rho_s$.
}
\end{figure}
{\it ASF-insulator quantum phase transitions}.--
Next, we address  the quantum phase transitions between the ASF and 
the (trivially) insulating phases at $n=0$ and $n=2$.  While SF 
to insulator transitions are often continuous, they can be driven first-order for attractive interactions~\cite{kuklov04b,radzihovsky08,lee10}. 
We find from our QMC simulations that the  
transition from the empty system ($n=0$) to the ASF is indeed first-order over an extended parameter regime. 
This is evident from  robust discontinuities in both the filling 
$n$ and superfluid density $\rho_s$, such as those shown in the left 
panel of Fig.~4 at $\mu/ |U|=-1.2$. 
From Ref.~\cite{lee10}, the transition is  
expected to turn continuous beyond a tricritical point near $\mu/|U|\approx -0.9$.
Performing simulations down to $\mu/|U|=-1.5$, we did not obtain evidence for a 
continuous transition, indicating  strong effects of the Feshbach resonance coupling for $\mu/|U|<-0.5$. 
The scenario of Ref.~\cite{lee10}, including a tricritical point, is  found to be 
realized for the transition from the ASF to the full system ($n=2$), which
we find to be
continuous beyond $\mu/|U|\approx -0.3$, cf. Fig.~1. 
The right panel of Fig.~4 shows the filling $n$ and the superfluid density $\rho_s$ along the line $\mu=0$, where no discontinuities are observed. 
To assess if this transition corresponds to that of a dilute bose gas of holes~\cite{sachdev99b},
we analyze the behavior close to the transition point in more detail. In that case and for 
two dimensions, the density $(2-n)$ of the holes 
should exhibit a linear increase $(2-n)=a(\tau-\tau_c) \ln(b/(\tau-\tau_c))$ with a logarithmic correction~\cite{zhitomirsky98}
in the vicinity of the transition point at $\tau_c=0.125$, while $\rho_s=(2-n)$ in the dilute limit~\cite{bernardet02}. 
As seen from Fig.~4 (right panel), both observables fit well to these functional forms;
the dynamical critical exponent equals $z=2$ at this transition~\cite{sachdev99b}, while our QMC temperature scales are sufficiently low to sample the finite system's ground state. 


{\it Conclusion}.--
The DSF in the considered boson lattice model
is restricted  to a narrow region of parameter space;  the anomalous jump in the stiffness at the BKT transition
should however provide a direct experimental signature of this phase.
The ASF-insulator transitions are found in overall agreement with the effective potential approach~\cite{lee10}, which 
however
underestimates the Feshbach resonance coupling effects at low filling.  
We established the presence of a tricritical point along the ASF to insulator transition line; determining its precise location and  critical properties remain  challenges for future studies. It will also be interesting to explore finite temperature transitions between the DSF and the ASF.

\begin{acknowledgements}
We thank H.P. B\"uchler, A. J. Daley, A.M. L\"auchli, S. Manmana, L. Pollet, 
K.P. Schmidt, and 
B. Svistunov
for discussions, and
acknowledge the allocation of CPU time at HLRS Stuttgart 
and NIC J\"ulich.
Support was also provided through 
the Studienstiftung des Deutschen Volkes (LB)
and
the DFG within SFB/TRR 21 (SW).

{\it Note added}.--
After the completion of our simulations, we became aware of  recent results~\cite{ng11} on the thermal phase transitions in an extended three-body constrained boson lattice model, consistent with our findings.

\end{acknowledgements}
\bibliographystyle{/user/home/people/wessel/tex/styles/prsty.bst}

\end{document}